\documentclass[final,5p,times,twocolumn]{elsarticle} 

\usepackage{lineno,hyperref}
\usepackage{siunitx}
\usepackage{hyperref}
\usepackage{mathtools}
\usepackage{rotating}
\usepackage{multirow}
\usepackage{xcolor,colortbl}
\usepackage{csquotes}
\usepackage{soul}
\usepackage{wrapfig}
\usepackage{amsmath}

\modulolinenumbers[5]

\journal{Nuclear Instruments and Methods in Physics Research Section A, }

\begin{document}

\begin{frontmatter}

\title{Particle Physics Readout Electronics and Novel Detector Technologies for
Neutron Science}


\author[a]{Thomas Block}
\author[a]{Markus Gruber}
\author[a]{Saime Gurbuz}
\author[a]{Jochen Kaminski}
\author[a]{Michael Lupberger\corref{mycorrespondingauthor}}
\cortext[mycorrespondingauthor]{Corresponding author}\ead{lupberger@physik.uni-bonn.de}
\author[a]{Divya Pal}
\author[a]{ Laura Rodriguez Gomez}
\author[a]{Patrick Schwaebig}
\author[a]{Klaus Desch}

\address[a]{Universit\"at Bonn, Bonn, Germany}

\begin{abstract}
Traditional thermal neutron detectors are based on Helium-3 as conversion and detection material due to its large neutron cross-section.
In light of the upgrade and construction of several neutron scattering facilities such as the European Spallation Source (ESS) and a simultaneous shortage of Helium-3, new detection technologies have been introduced. The most prominent one is to use solid converts with a large thermal neutron cross-section such as Gadolinium and Boron. Those materials emit charged particles when hit by a neutron. The technique then relies on detection and/or tracking of the charged particle, as in detectors of particle physics. At the same time, this requires an increase of the readout channels by an order of magnitude with the advantage of also increasing the position resolution by the same amount compared to traditional neutron detectors. A prime example is the Gadolinium Gas Electron Multiplier (GdGEM) detector for the NMX instrument at ESS jointly developed by the CERN Gaseous Detector Group and the ESS Detector Group.

In this contribution, some of our efforts to transfer particle physics detectors and readout electronics to neutron science will be presented.
We employed the VMM3a chip, originally designed for the ATLAS New Small Wheel upgrade, to read out a GEM-based neutron detector. The Timepix3 chip is employed in a neutron Time Projection Chamber as well as to read out a neutron-sensitive Micro-Channel Plate detector. Those readout chips are integrated into the Scalable Readout System of the RD51 collaboration.
\end{abstract}


\end{frontmatter}


\section{Introduction}\label{sec:Intro}
\begin{wrapfigure}{l}{0.5\linewidth}
\vspace{-4.0mm}
    \centering
    \includegraphics[width=\linewidth]{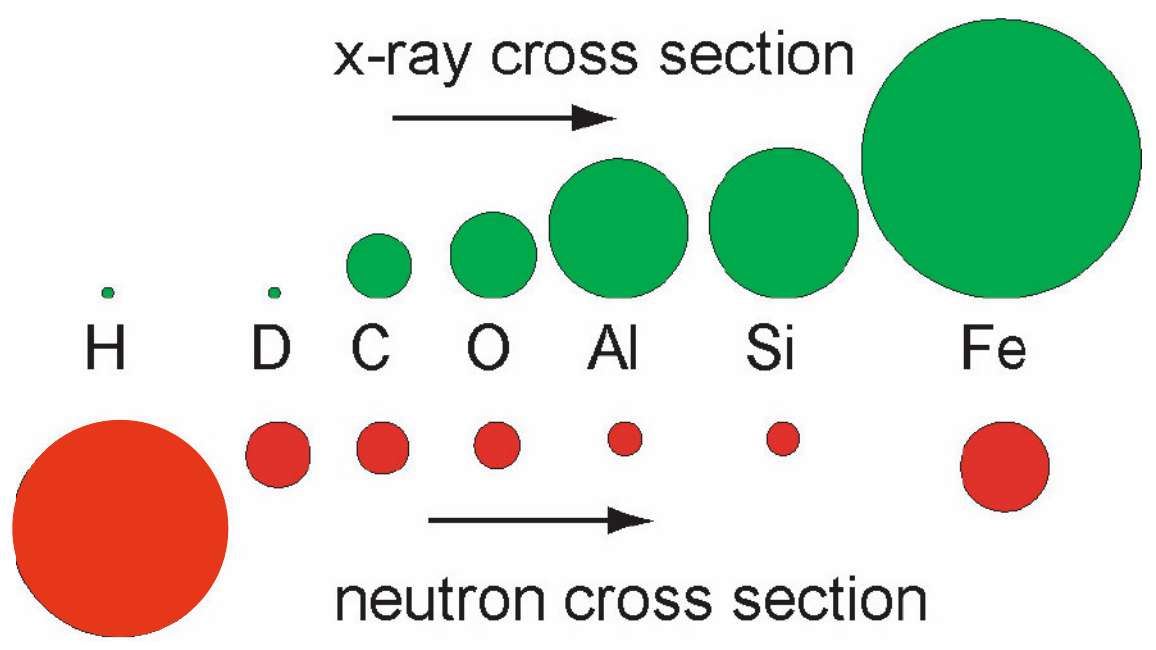}
    \vspace{-8.0mm}
    \caption{Comparison of X-ray and neutron cross-section for certain elements, from \cite{article}.}
    \label{fig:nXsection}
\vspace{-2.0mm}
\end{wrapfigure}

\vspace{-1.0mm}
Neutrons are complementary probes to investigate the composition and structure of objects. In contrast to X-rays, the absorption and scattering cross-section does not increase with the atom size but varies significantly over the element spectrum. This is due to the fact that X-rays interact with the atomic shell, while neutral neutrons interact with the atom core e.g. through capture. Fig.~\ref{fig:nXsection} compares the X-ray and neutron cross-section for several elements. In particular hydrogen, as the lightest element, interacts strongly with neutrons and weakly with X-rays, while for iron, the situation is the opposite. 

\begin{figure}[htbp]
\vspace{-2.0mm}
    \centering
    \includegraphics[width=0.9\linewidth]{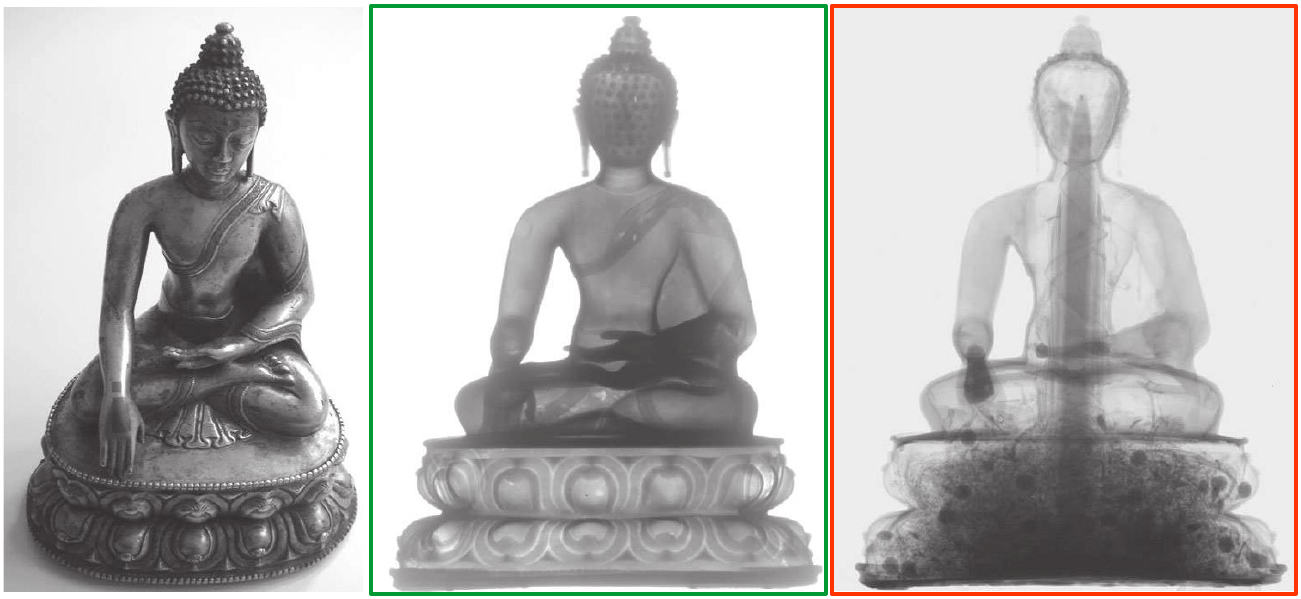}
    \vspace{-3.0mm}
    \caption{Optical (left), X-ray (middle) and neutron (right) imaging of a Buddha statue, from \cite{lehmann2010investigation}.}
    \label{fig:Buddha}
\vspace{-2.0mm}
\end{figure}

\begin{figure}[htbp]
\vspace{-4.0mm}
    \centering
    \includegraphics[width=1.0\linewidth]{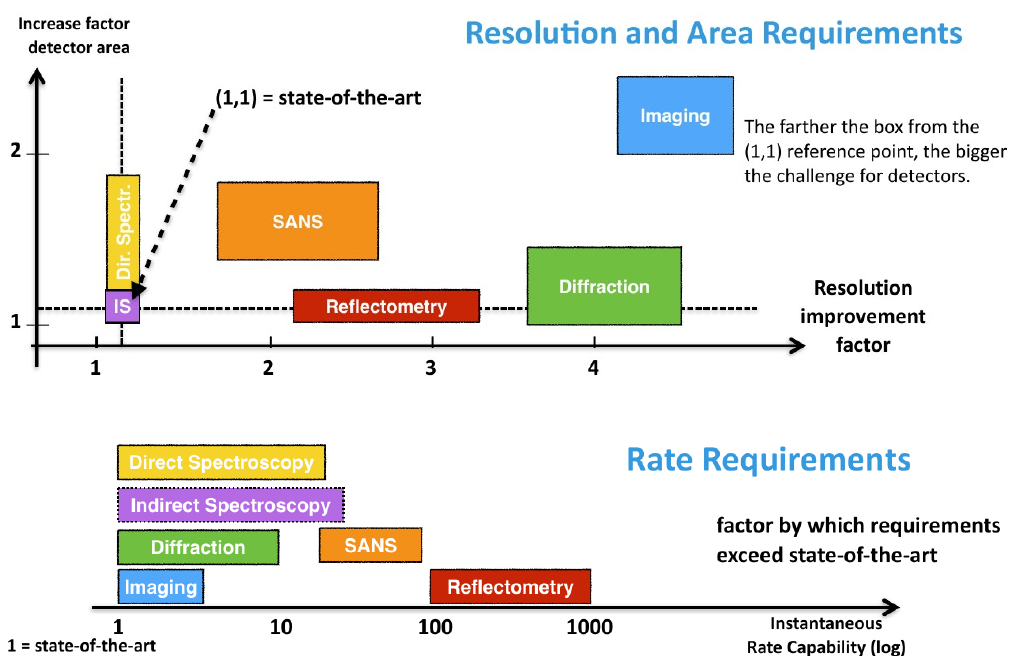}
    \vspace{-7.0mm}
    \caption{Main requirements of detectors for the ESS, from \cite{RichardCERNdetectorSeminar}.}
    \label{fig:ESSdet}
\vspace{-3.0mm}
\end{figure}

An advantage of this opposition can be taken when investigating objects with X-rays and neutrons, as shown in Fig.~\ref{fig:Buddha}. A Buddha statue (optical image on the left) reveals its metallic composition in X-ray imaging (middle image) and internal structure in neutron imaging (right image), in this case, a wooden stick and some other organic support giving hints how the statue was built. Other applications of neutron probes are to determine the position of hydrogen atoms in pharmaceutical drugs, the detection of water (e.g. to localise cracks in bridges or determine the soil moisture) or to investigate organic materials, to list only a few.
The systematic detection of, in our case thermal, neutrons in scientific experiments relies on their conversion to charged particles. For that reason, elements with a high neutron cross-section, which simultaneously emit a charged particle, are key. Helium-3, with the process
\begin{equation}
    {^3\text{He}} + \text{n} \rightarrow {^3\text{H}^{+}} + \text{p}^+
\end{equation}
is an ideal detection medium. Both, the emitted proton and tritium ionise the gas medium and thus produce seeds for detection as in particle physics gaseous detectors. Helium-3 tubes are usually operated as Geiger counters and can be arranged in an array to detect e.g. neutron diffraction patterns off an object under investigation in a neutron beam. The position resolution of such devices is limited to the lower \SI{}{\milli\meter} regime, the rate capability to $\mathcal{O}$(\SI{30}{\kilo\hertz}), while the detection efficiency of the (high-pressure) $\approx$~\SI{1}{\centi\meter} diameter tubes is close to \SI{100}{\percent}.
New generation neutron sources such as the European Spallation Source (ESS) have much more stringent requirements for their detectors as can be seen in Fig.~\ref{fig:ESSdet}. The detector area and resolution have to be increased moderately, while many applications require orders of magnitude rate capability improvements. At the same time, the global Helium-3 availability is difficult. This is due to the fact that on the one hand, the demand for Helium-3 detectors for homeland security has increased\footnote{E.g. to detect uranium, which has a very high neutron cross-section.} and on the other hand, Helium-3 production, as a decay product of tritium in the reduced thermonuclear weapon stockpile, is reduced.
This led to the development of novel neutron detectors based on solid converters. The requirements for such converter materials are a high neutron-cross section, the emittance of charged particles and that they are easy to handle. The secondary charged particle then allows the application of state-of-the-art particle physics tracking detectors, such as gaseous detectors. Gadolinium, with one of the largest cross-sections, is a prominent candidate material. However, the emitted \SI{}{\mega\electronvolt} electron ionisation track cannot be distinguished from Compton-scattering background photons which usually come with a neutron beam unless specifically suppressed. The NMX instrument at ESS applies such a detector called GdGEM (Gadolinium Gas Electron Multiplier)~\cite{pfeiffer2016first,lupberger2020srs} and is a prime example of technology transfer from particle physics to neutron science.

Boron-10 is another element suitable for neutron detection. Its cross-section is comparable to Helium-3 and it emits two charged particles in the process
\begin{equation}
    {^{10}\text{B}} + \text{n} \rightarrow {^7\text{Li}^{+++}} + {^4\text{He}^{++}} 
\end{equation}

An issue of solid converters is that the secondary particles need to leave the material. This only allows for thin converter layers of a few \SI{}{\micro\meter}. At this thickness, Gadolinium absorbs \SI{100}{\percent} of thermal neutrons, Boron only converts about \SI{5}{\percent}. For that reason, boron-based neutron detectors require several layers. In the following, we present three particle physics concepts and readout electronics investigated for novel neutron detectors.
\section{Micro-Channel Plate detector with Timepix3 readout }\label{sec:nMCP}
\begin{figure}[htbp]
\vspace{-4.0mm}
    \centering
    \includegraphics[width=0.7\linewidth]{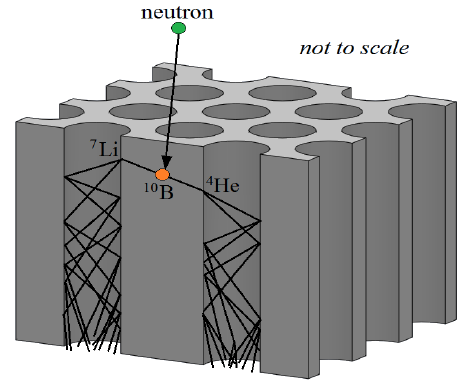}
    \vspace{-3.0mm}
    \caption{Basic working principle of a boronated MCP, from~\cite{craft2013afip}.}
    \label{fig:nMCP_working}
\vspace{-3.0mm}
\end{figure}

A Micro-Channel Plate (MCP) is a high-resistive material e.g. glass, with fine pores and electrodes on its top and bottom side to apply a high potential difference, see Fig.~\ref{fig:nMCP_working}. It is operated in a vacuum. The walls of the pores are coated with a semiconductor material. The MCP we use is additionally doped with Boron-10\footnote{It is also doped with Gadolinium and a single layer provides a \SI{50}{\percent} detection efficiency for thermal neutrons.}. The pore pitch is low enough to allow at least one of the conversion fragments of a neutron interaction to reach the semiconductor walls, where it extracts electrons that enter the micro channel. Those electrons are accelerated and hit the walls again, which generates an electron avalanche. Placing a charge-sensitive device below the MCP allows for charge collection i.e. neutron detection.
This detector type provides a very high spatial resolution down to a few \SI{}{\micro\meter} when a highly granular pixel chip is employed as a charge-sensitive readout. \cite{tremsin2012high} used the Timepix, where the shutter-based readout led to large dead times. In our approach, we employ the Timepix3, which we implemented in the Scalable Readout System of the RD51 Collaboration~\cite{Gruber_2022}, with continuous, self-triggered data-driven readout. With this chip, we plan for a quad Timepix3 readout, the detector could cope with rates in the \SI{10}{\mega\hertz} regime with $\mathcal{O}$(\SI{100}{\nano\second}) time resolution on a small area of a few \SI{}{\square\centi\meter}.
\begin{figure}[htbp]
\vspace{-4.0mm}
    \centering
    \includegraphics[width=1.0\linewidth]{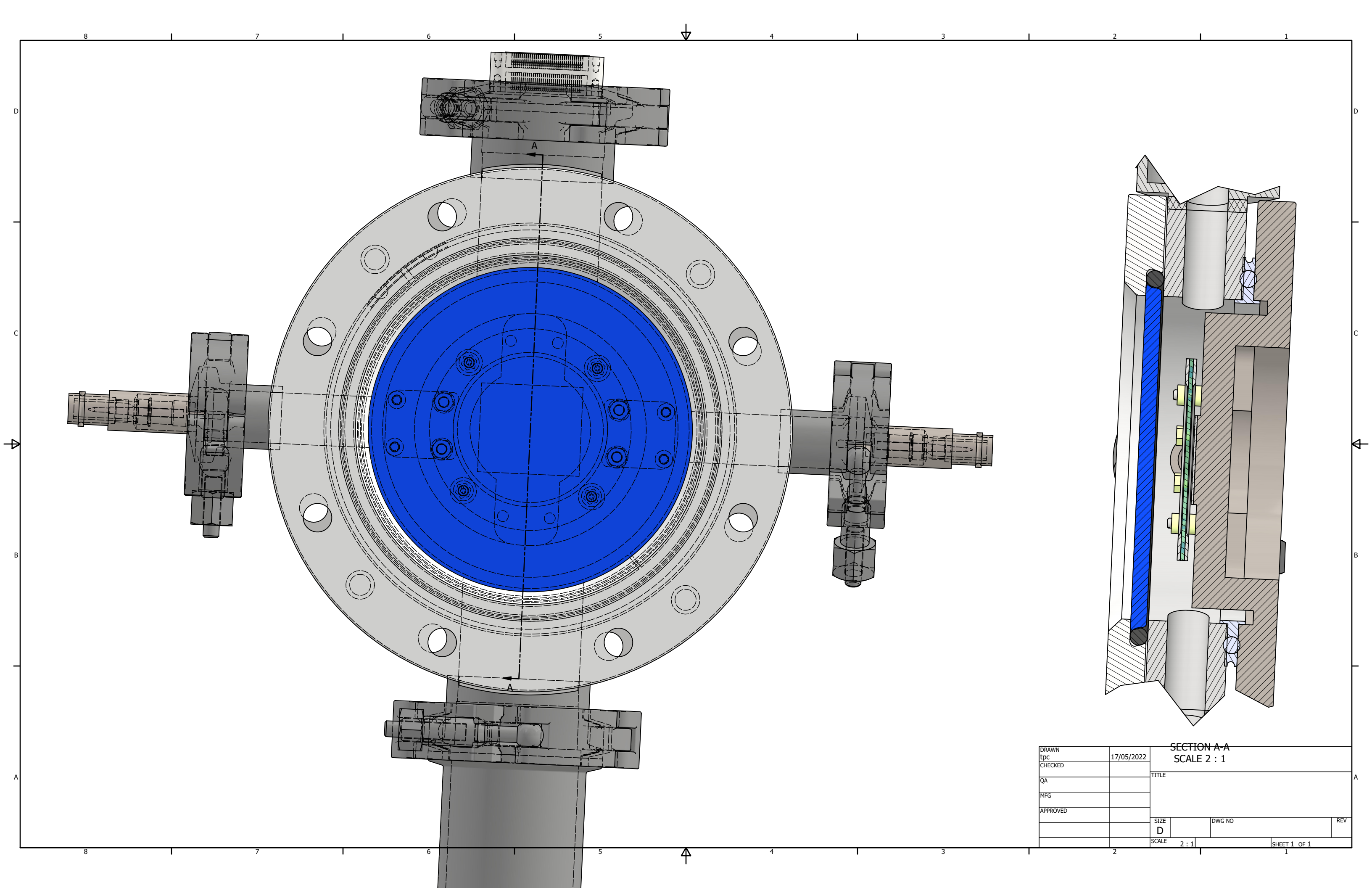}
    \vspace{-3.0mm}
    \caption{Technical drawing of the neutron MCP detector with a view through the quartz/fused silica view-port (left) and a profile view (right), with window (blue), MCP (green) and Timepix3 (dark gray directly below the MCP).}
    \label{fig:nMCP_layout}
\vspace{-3.0mm}
\end{figure}

\begin{wrapfigure}{l}{0.5\linewidth}
\vspace{-0.0mm}
    \centering
    \includegraphics[width=1.0\linewidth]{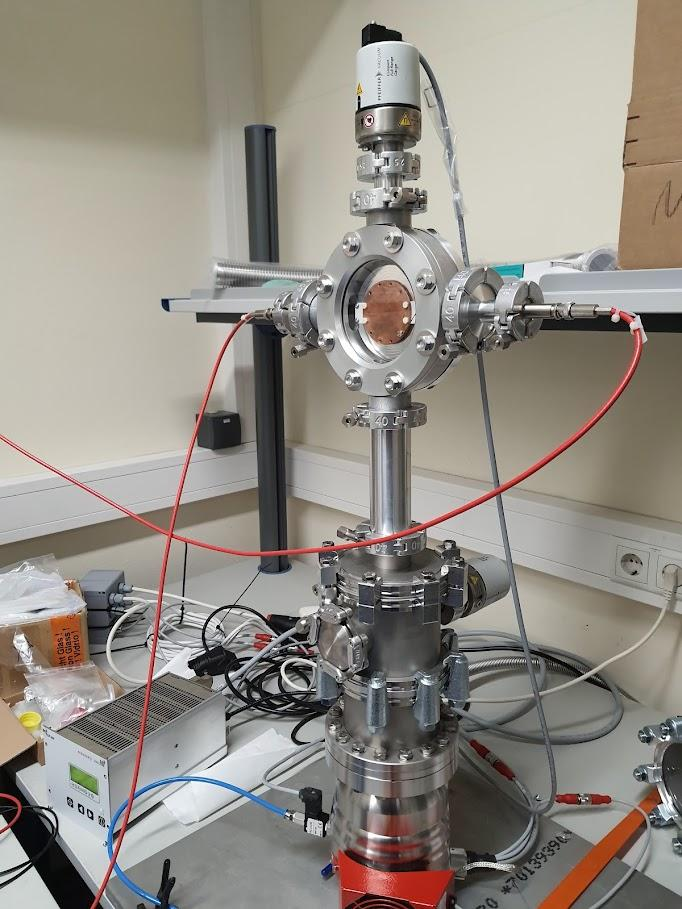}
    \caption{Picture of the neutron MPC detector setup with the vacuum system in the laboratory.}
    \label{fig:nMCP_lab}
\vspace{-4.0mm}
\end{wrapfigure}
Fig.~\ref{fig:nMCP_layout} shows a technical drawing of the neutron MCP detector. It is housed in an aluminium (necessary for low neutron activation) vacuum vessel with high-voltage feedthroughs enclosed by a neutron-transparent quartz window. The MCP and Timepix3 are only separated by a thin gap to mitigate dispersion. The chip(s) will be placed on a cooling unit serving as back-side flange with a feedthroughs for a connector. All other electronics is placed outside the vacuum. Fig.~\ref{fig:nMCP_lab} shows the setup, consisting mainly of the vacuum system, in the laboratory, while the high-voltage stability was tested with a copper plate instead of the MCP. The vacuum was also tested to reach the required \SI{1e-6}{\milli\bar}.

\section{Time-Projection Chamber detector with GridPix readout}\label{sec:nTPC}
\begin{figure}[htbp]
\vspace{-2mm}
    \centering
    \includegraphics[width=1.0\linewidth]{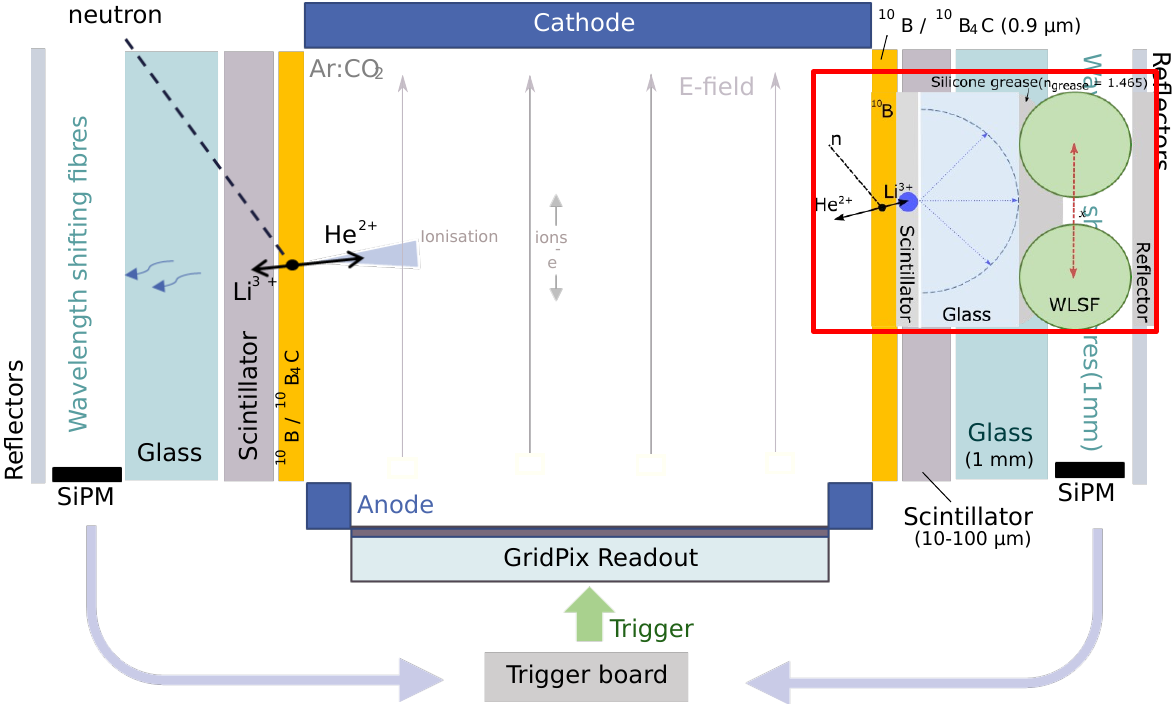}
    \caption{Schematic layout of the neutron TPC. The inlay picture shows the neutron conversion in the boronated walls and the two-particle process required for the triggering.}
    \vspace{-4mm}
    \label{fig:nTPC_layout}
\vspace{-0mm}
\end{figure}
A Time-Projection Chamber (TPC) is a gas-filled volume with a cathode and an anode to induce an electric drift field. Charged particles traversing the volume generate an ionisation path, of which the electrons are drifted towards the anode for detection. In our setup, we employ a highly segmented anode with intrinsic gas amplification as readout, the GridPix technology~\cite{COLAS2004506,Kaminski:2017bgj}. It consists of a Timepix or Timepix3 chip with a resistive protection layer and a post-processed Micromegas grid on top. In a photo-lithographic process, the grid holes are aligned one to one with the charge collecting bump-bond pads of the chip. Single electrons from the drift region above are amplified and the avalanche induces a signal only in the single pixel below, which enables single-electron detection. In a TPC, those individual electrons are the projection of the ionisation path of the traversing particle. Recording the arrival time at the readout, in addition, allows a relative three-dimensional track reconstruction.

Fig.~\ref{fig:nTPC_layout} shows a schematic drawing of the neutron TPC. In addition to the traditional layout, the side walls are coated with Boron-10, such that one of the neutron conversion fragments can enter the gas volume and generate the ionisation path. The second fragment is used to generate a time reference trigger. The red box inlay figure in Fig.~\ref{fig:nTPC_layout} shows a detailed view of the TPC wall. It consists of a glass plate with a scintillator and a Boron-10 coating. The light induced by the conversion fragment is coupled to WaveLength Shifting Fibers (WLSFs) and guided to Silicon Photo-Multipliers (SiPMs), which provide a trigger signal to the Timepix/Timepix3 readout as a time reference for absolute three-dimensional track and neutron impact point reconstruction down to a few \SI{10}{\micro\meter} at $\mathcal{O}$(\SI{100}{\kilo\hertz}) rates. 
\begin{figure}[htbp]
\vspace{-4mm}
    \centering
    \includegraphics[width=0.7\linewidth]{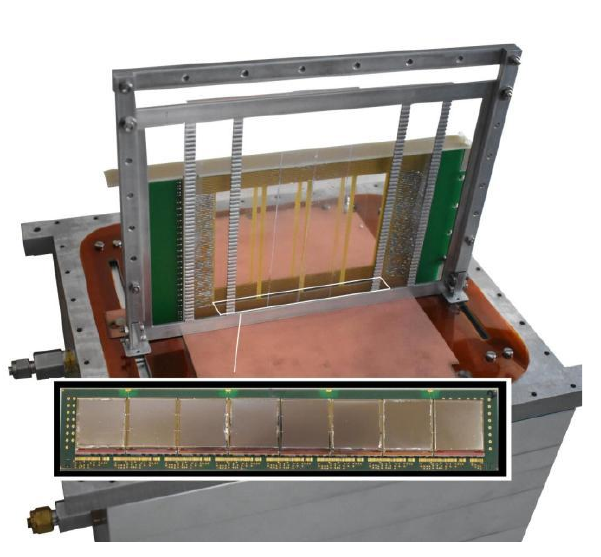}
    \vspace{-4mm}
    \caption{Neutron TPC with support frame for the scintillator bars, cathode, wire field cage and anode plate. The sensitive readout area consists of an array of eight Timepix/Timepix3 chips (inlay picture).}
    \label{fig:n_TPC_lab}
\vspace{0mm}
\end{figure}

As the TPC walls ( $\approx 11 \times 8$~\SI{}{\square\centi\meter} active area) are used for neutron conversion, field shaping is implemented by wires, for which the holding structure is shown in Fig.~\ref{fig:n_TPC_lab} together with other components of the detector. Placing several detectors next to each other allows for several layers to achieve a high detection efficiency.

In a primary test with cosmic particles and external scintillator triggers, charge particle tracks were recorded by the readout array consisting of eight Timepix GridPix chips shown in Fig.~\ref{fig:nTPC_tracks}.
\textbf{\begin{figure}[htbp]
\vspace{-6.0mm}
    \centering
    \includegraphics[width=1.0\linewidth]{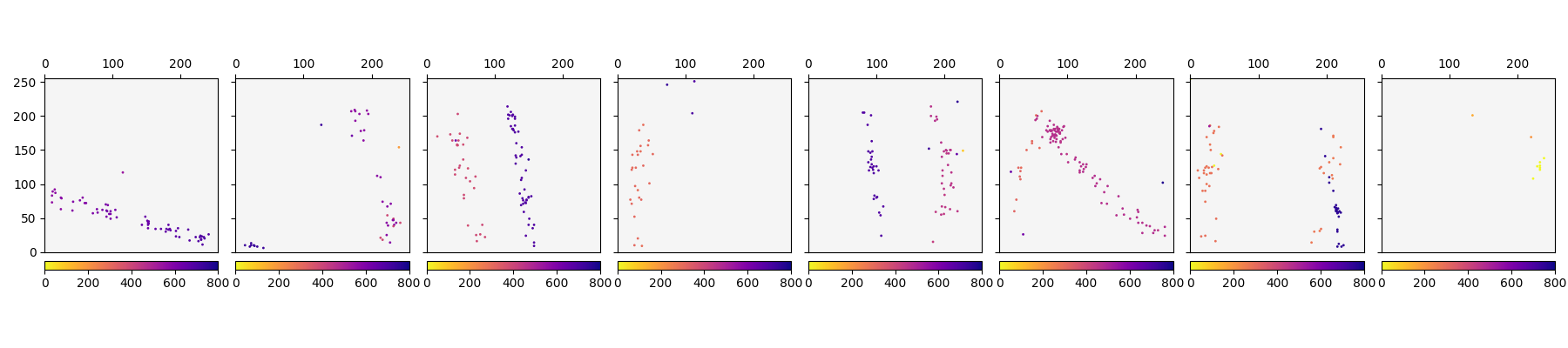}
    \vspace{-4.0mm}
    \caption{Tracks of cosmic particles recorded with the TPC without boron-coated side walls and an external scintillator trigger. The colour code on the z-axis marks the arrival time.}
    \label{fig:nTPC_tracks}
\vspace{-4.0mm}
\end{figure}}
\section{Gas Electron Multiplier (GEM) detector with VMM3a readout
}\label{sec:nGEM}
Gas Electron Multipliers (GEMs) are one of the main gaseous detector types. They are made of an insulating foil typically of \SI{50}{\micro\meter} thickness with copper electrodes on both sides. The structure is perforated by holes of $\approx$ \SI{50}{\micro\meter} diameter with $\approx$~\SI{70}{\micro\meter} pitch. It is operated in a gas volume with a high voltage applied between the two sides, such that gas amplification occurs in the holes when an electron enters. When the GEM top side is coated with Boron, the highly ionising neutron conversion fragments (ions) generate a large number of electrons. Those drift into adjacent GEM holes and are amplified. A single GEM foil is sufficient to induce a signal strength on a separated readout anode measurable with recent front-end electronics. Such a concept is employed in the CASCADE neutron detector~\cite{kohli2016cascade}. High detection efficiency is achieved by a Boron-coated cathode and several layers of Boron-coated GEMs. 

\begin{figure}[htbp]
\vspace{-4.0mm}
    \centering
    \includegraphics[width=0.7\linewidth]{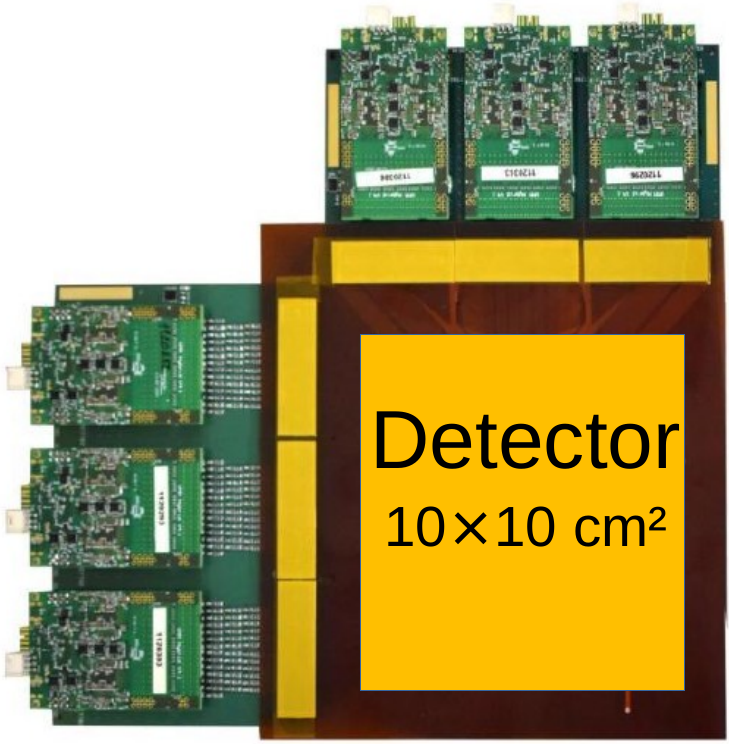}
    \vspace{-3.0mm}
    \caption{Prototype neutron GEM detector single layer with GEM foil and VMM3a front-end electronics.}
    \label{fig:nGEM}
\vspace{-8.0mm}
\end{figure}

In our concept, we read out the signals from the GEMs directly. Fig.~\ref{fig:nGEM} shows a $10 \times 10$~\SI{}{\square\centi\meter} GEM (in this case without Boron) and the front-end electronics currently under test. We apply the RD51 SRS with the VMM3a as front-end chip~\cite{SRS_VMM}. The final detector foresees about eight larger layers, with a total of about 100 front-end boards, which requires dedicated front-end cooling. The detector will cover a large sensitive area of $30 \times 30$~\SI{}{\square\centi\meter} and should cope with high rates $\mathcal{O}$(\SI{10}{\mega\hertz}) at medium $\mathcal{O}$(\SI{100}{\micro\meter}) spatial resolution. 
\section{Conclusion}\label{sec:Conc}
 Next-generation neutron detectors use solid converters emitting charged particles after neutron capture. This allows using existing particle physics concepts to build large area detectors. They can detect and track those secondary particles and allow the neutron impact point reconstruction with high spatial resolution. Using state-of-the-art particle physics readout electronics, readout rates, time and space resolution superior to previous neutron detectors can be reached. In our approaches, we transfer well-established particle physics readout electronics and detector concepts to neutron science to build novel thermal neutron detectors with different capabilities. We combine a Boron-doped MCP with the Timepix3 chip, a GridPix TPC with Boron-coated walls and Boron-coated GEMs with the VMM3a readout chip. All detectors are under test and we apply the RD51 SRS, where we implement the Timepix3, as a readout platform.
\section{Acknowledgements}\label{sec:Ack}
We would like to thank the organisers of the 15th Pisa Meeting on Advanced Detectors for the perfect conference and the possibility to present our work in a presentation.
We also recognise the contributions of Dr. Markus Köhli from the University of Heidelberg for his continued assistance, advice, ideas and support within this project.
Parts of this work have received funding from the European Union’s Horizon 2020 research and innovation programme under the Marie Sklodowska-Curie grant agreement no. 846674
as well as from the German Federal Ministry of Education and Research under grant no. 05K19PD1.

\bibliography{neutrons}

\begin{thebibliography}{10}
\expandafter\ifx\csname url\endcsname\relax
  \def\url#1{\texttt{#1}}\fi
\expandafter\ifx\csname urlprefix\endcsname\relax\def\urlprefix{URL }\fi
\expandafter\ifx\csname href\endcsname\relax
  \def\href#1#2{#2} \def\path#1{#1}\fi

\bibitem{article}
K.~Kihm, E.~Kirchoff, M.~Golden, J.~Rosenfeld, S.~Rawal, D.~Pratt, A.~Swanson,
  H.~Bilheux, L.~Walker, S.~Voisin, D.~Hussey, D.~Jacobson, Neutron imaging of
  alkali metal heat pipes, Physics Procedia 43 (2013) 323–330.
\newblock \href {https://doi.org/10.1016/j.phpro.2013.03.038}
  {\path{doi:10.1016/j.phpro.2013.03.038}}.

\bibitem{lehmann2010investigation}
E.~H. Lehmann, S.~Hartmann, M.~O. Speidel, Investigation of the content of
  ancient tibetan metallic buddha statues by means of neutron imaging methods,
  Archaeometry 52~(3) (2010) 416--428.

\bibitem{RichardCERNdetectorSeminar}
R.~Hall-Wilton, \href{https://indico.cern.ch/event/979864/}{Detectors for
  neutron scattering science at the european spallation source}, in:
  Presentation given at the CERN detector seminar, 2020.
\newline\urlprefix\url{https://indico.cern.ch/event/979864/}

\bibitem{pfeiffer2016first}
D.~Pfeiffer, F.~Resnati, J.~Birch, M.~Etxegarai, R.~Hall-Wilton,
  C.~H{\"o}glund, L.~Hultman, I.~Llamas-Jansa, E.~Oliveri, E.~Oksanen, et~al.,
  {First measurements with new high-resolution gadolinium-GEM neutron
  detectors}, JINST 11~(05) (2016) P05011.

\bibitem{lupberger2020srs}
M.~Lupberger, F.~M. Brunbauer, Y.~Huang, H.~M{\"u}ller, E.~Oliveri,
  D.~Pfeiffer, L.~Ropelewski, M.~Van~Stenis, P.~Thuiner, {SRS VMM readout for
  Gadolinium GEM-based detector prototypes for the NMX instrument at ESS}, in:
  Journal of Physics: Conference Series, Vol. 1498, IOP Publishing, 2020, p.
  012050.

\bibitem{craft2013afip}
A.~Craft, W.~Williams, M.~Abir, D.~Wachs, Afip-7 tomography--2013 status
  report, Tech. rep., Idaho National Lab.(INL), Idaho Falls, ID (United States)
  (2013).

\bibitem{tremsin2012high}
A.~S. Tremsin, J.~V. Vallerga, J.~B. McPhate, O.~H. Siegmund, R.~Raffanti,
  {High Resolution Photon Counting With MCP-Timepix Quad Parallel Readout
  Operating at \textgreater~1~kHz Frame Rates}, IEEE TNS 60~(2) (2012)
  578--585.

\bibitem{Gruber_2022}
M.~Gruber, K.~Desch, T.~Hemperek, J.~Kaminski, L.~Richarz, T.~Schiffer,
  {{SRS}-based Timepix3 readout system}, JINST 17~(04) (2022) C04015.
\newblock \href {https://doi.org/10.1088/1748-0221/17/04/c04015}
  {\path{doi:10.1088/1748-0221/17/04/c04015}}.

\bibitem{COLAS2004506}
P.~Colas, A.~Colijn, A.~Fornaini, Y.~Giomataris, H.~{van der Graaf}, E.~Heijne,
  X.~Llopart, J.~Schmitz, J.~Timmermans, J.~Visschers, {The readout of a GEM or
  Micromegas-equipped TPC by means of the Medipix2 CMOS sensor as direct
  anode}, NIM A 535~(1) (2004) 506--510.
\newblock \href {https://doi.org/https://doi.org/10.1016/j.nima.2004.07.180}
  {\path{doi:https://doi.org/10.1016/j.nima.2004.07.180}}.

\bibitem{Kaminski:2017bgj}
J.~Kaminski, Y.~Bilevych, K.~Desch, C.~Krieger, M.~Lupberger, {GridPix
  detectors \textendash{} introduction and applications}, NIM A 845 (2017)
  233--235.
\newblock \href {https://doi.org/10.1016/j.nima.2016.05.134}
  {\path{doi:10.1016/j.nima.2016.05.134}}.

\bibitem{kohli2016cascade}
M.~K{\"o}hli, M.~Klein, F.~Allmendinger, A.~Perrevoort, T.~Schr{\"o}der,
  N.~Martin, C.~Schmidt, U.~Schmidt, Cascade-a multi-layer boron-10 neutron
  detection system, in: Journal of Physics: Conference Series, Vol. 746, IOP
  Publishing, 2016, p. 012003.

\bibitem{SRS_VMM}
M.~Lupberger, L.~Bartels, F.~Brunbauer, M.~Guth, S.~Martoiu, H.~Müller,
  E.~Oliveri, D.~Pfeiffer, L.~Ropelewski, A.~Rusu, P.~Thuiner, {Implementation
  of the VMM ASIC in the Scalable Readout System}, NIM A 903 (2018) 91 -- 98.
\newblock \href {https://doi.org/https://doi.org/10.1016/j.nima.2018.06.046}
  {\path{doi:https://doi.org/10.1016/j.nima.2018.06.046}}.

\end{thebibliography}

\end{document}